\documentclass[]{aa}
\usepackage{txfonts, graphicx, epsfig, natbib,url} 
\bibpunct{(}{)}{;}{a}{}{,} 
\usepackage{multirow}
\begin{document}

\twocolumn{%
%\onecolumn{%

  \title {Digitized archive of the Kodaikanal images: Representative results of solar cycle variation from sunspot area determination }
 
 \author { B. {Ravindra}, 
              T. G.  {Priya}, K.  {Amareswari}, M.  {Priyal},  A. A.  {Nazia},  D.  {Banerjee} }

         \institute{ Indian Institute of Astrophysics,
         Koramangala,
         Bengaluru-560 034, India}
\offprints{D. Banerjee \email{dipu@iiap.res.in}}
\date{\today}     

\begin{abstract}
%Context
{ Sunspots  have been observed since Galelio Gelili invented the telescope. Later, sunspot drawings have been upgraded to 
image storage using photographic plate in the second half of nineteenth century. These photographic images 
are valuable data resources for studying long-term changes in the solar magnetic field and its influence on the Earth's climate and weather. }
% Aims: 
{Digitized photographic plates cannot be used directly for the scientific analysis. 
It requires certain steps of calibration and processing before using them for extracting any 
useful information. The final data can be used to study solar cycle variations over several cycles.}
%{\it Methods:} 
{We digitized more than 100 years of white-light images stored in photographic 
plates and films that are available at Kodaikanal observatory starting from 1904. The images 
were digitized using a 4k$\times$4k format CCD-camera-based digitizer unit.The digitized images were  calibrated for relative plate density and aligned in such a way that the solar north is in upward 
direction. A semi-automated  sunspot detection 
technique was used to identify the sunspots on the digitized images. }
%{\it Results:} 
{In addition to  describing the calibration procedure and availability of the data, we here present preliminary results on the sunspot area measurements and their variation with time.  The results show that
the white-light images have a uniform spatial resolution throughout the 90 years of observations.
However, the contrast of the images decreases from 1968 onwards. The images are circular and do not
show any major geometrical distortions. The measured monthly averaged  sunspot areas  closely match 
the Greenwich sunspot area over the four solar cycles studied here. The yearly averaged sunspot area shows a
high degree of correlation with the Greenwich sunspot area. Though the monthly averaged sunspot number 
shows a good correlation with the monthly averaged sunspot areas, there is a slight anti-correlation 
between the two during solar maximum.  }
%{\it Conclusions:}
{The Kodaikanal data archive is hosted at \url{http://kso.iiap.res.in}. The long time sequence of the Kodaikanal 
white-light images provides a consistent data 
set for sunspot areas and other  proxies. Many studies  can be performed using Kodaikanal data alone without requiring intercalibration between different data sources.  }
\end{abstract}

\titlerunning{Kodaikanal Digitized Archive}
\authorrunning{B. Ravindra}

\maketitle

\section{Introduction}
\label{sec:introduction}
For about the past 100 years, many solar observatories around the world have taken the Sun's images  on photographic plates. Generally, there is a high interest in these data sets simply because 
they provide the history of the Sun's activity. However, there are some limitations in using these
data sets in their current format. This is mainly because it is very cumbersome to extract any 
information from its current format. Hence, in many observatories the digitization of the
photographic plates has been carried out.

Mt. Wilson observatory has 70 years of Ca~II~K data sets and similar intervals of white-light data 
sets. The Arcetri observatory, the Big-Bear Solar Observatory (BBSO), and the Kitt-Peak solar observatory 
also have 20-40 years of Ca~II~K data sets taken on photographic plates. All these data sets 
have been digitized in the past ten to twenty years. Kodaikanal observatory, which was established in 1899 \citep{hasan2010}, has a history of taking solar images in Ca~II~K, H$_{\alpha}$, and white-light in
photographic plates. In all these three data sets white-light data have been obtained from the same solar telescope without changing any of its optics  for more than
one hundred years, and it is being continued  until today. This also provides some  overlap between the old and new data taken at Kodaikanal using a different set of telescopes.  Since the same telescope
has been used to acquire the white-light images, the data are extremely useful for carrying out
the studies of long-term activity on the Sun. In an earlier attempt, the digitization of the Kodaikanal white-light data was carried out by \cite{sivaraman1993}. In this digitization of photographic plates
a CalComp digitizer was used. This digitizer acquires the position of the
solar limb and the sunspot areas by using the cross hair that is kept on the limb and sunspots. 
However, these authors have not stored the images of the Sun in digitized format. To improve the spatial 
resolution of the digitized data and also to make them available to other researchers who may be 
interested in this century of data, we digitized the white-light images of the Sun using a large CCD camera. The digitized images are hosted at~\url{http://kso.iiap.res.in}. 

In this paper, we briefly describe the telescope used for obtaining the solar white-light images for the past hundred years, the content of the images, the period of data acquisition, the digitization of the century data, 
the data reduction, and the calibration procedures. We also present some preliminary results obtained 
by the digitized data archive of the white-light images.  The total area of sunspots detected on the solar disk is one of the fundamental proxies of solar magnetic activity \citep{2005A&A...443.1061B}. A complete, consistent and reliable time series of the sunspot area as determined from a single optical system is very useful.  We present our sunspot area results for four cycles and compare them with the Greenwich results. 
\section{Century-old telescope and solar photographs}
\label{sec:cot}
The white-light telescope consists of a 10-cm objective lens with f/15 beam that is capable of
producing 20.4 cm diameter images of the Sun after enlarging the primary image with the additional optics. 
Whenever the sky is clear, the images in white-light were captured on photographic plates since 1904. 
Later, in 1912, this objective lens was replaced with a Cooke photo-visual lens of the same size. The
final image size was the same as before. After taking the setup to Kashmir (India) for a while, 
the same setup was
reinstalled at Kodaikanal in 1917. Since June 13, 1918, the 15-cm achromatic lens has been used, which has 
a focal length of 240 cm. In the focal plane a green color filter was used as well, which 
improved the quality of the solar image. From 1918 to untill today the same telescope has been used 
for obtaining the white-light images of the Sun. More detailed descriptions of the telescope parts, 
its dismantling, reassembling, and modifications can be found in \cite{sivaraman1993}.

The telescope with the 15-cm objective is mounted in equatorial configuration.
In the beginning of the observations a cross wire of solar disk size was exposed across the
plate that indicates the east-west and north-south position of the sky. This was 
continued until 1908; later on, the cross wire was replaced with a 
single thin wire of solar disk size  kept in the focal plane of the telescope. The position of the 
wire in the image represents the east-west direction in the sky on each image.

Starting from 1904, the solar white-light images have been stored in Lantern photographic-plates
of size 25.4 sq. cm. In 1975 these plates were decommissioned and hence in Jan 1976, they
were replaced by high-contrast film of size 25.4 cm $\times$ 30.5 cm. The density-to- 
intensity value bar codes were used since the year 1969. Before that these bar codes were not
available. The date and time of the observations were written in one corner of the emulsion side of the
photographic plate (see Fig.~\ref{fig:2}). The exposed photographic plates are kept in thick paper envelopes and were stored in cupboards at Kodaikanal observatory under dry conditions for many years. Most of the plates are still in a good condition, except for a few plates that acquired scratches, dusts, and fungus in a few places. 
A log book was maintained for each day of the observations that describes  the sky
conditions, as well as the date and time of the solar image acquisition.

\subsection{Observational data}
\label{sec:od}
The white-light observations of the Sun started in early 1904. The same observations are continued 
until today. The size of the solar image is about 20 cm in diameter. Normally, in 
the morning the Earth's atmospheric seeing is good at Kodaikanal. Clouds cover the sky in the afternoon, and  
in the evening it clears up again \citep{bappu1967,smith1895}.  Most of the time 
the images are taken early in the morning before 10~hr (IST). On most days one image is 
obtained and some days a few more images are taken, depending on the sky conditions. Once in a
month the overlap of solar image is taken by stopping the telescope tracking. One exposure is 
taken at the top and one at the bottom of the photographic plate. An intersection 
of these two images provides the north-south direction of the sky. However, these images are not 
available regularly to find the north-south direction of the sky. Hence we discard these images 
because we have an image of a straight line thin wire representing the east-west direction of the sky. 

\begin{figure}
\begin{center}
\includegraphics[width=0.45\textwidth]{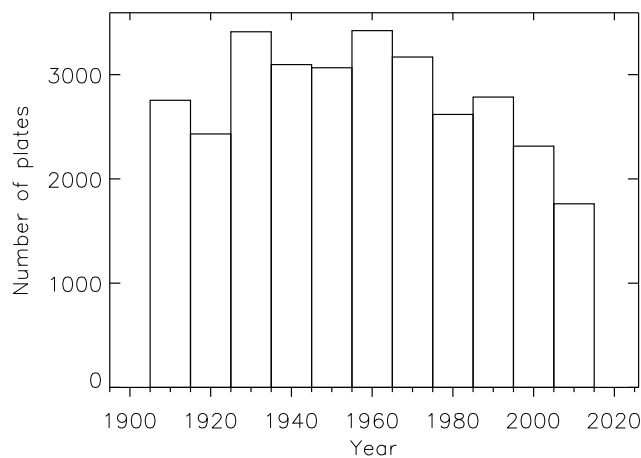}
\end{center}
\caption{Histogram of the number of white-light image plates available in the archive.} 
\label{fig:1}
\end{figure}

In 106 years of observations, 31800 plates covering over 31000 days were 
acquired. Figure \ref{fig:1} shows the histogram of the number of white-light observational images.
At first, only few images were acquired,
but many data were collected between 1930 to 1980.  The number of plates
increases during the period from 1960 to 1970. This is just after the announcement of the international
geophysical year (IGY). On average, about 290 white-light images were acquired each year. 

\section{Digitizer unit}
\label{sec:du}
The digitizer unit has a uniform source sphere of 1-m diameter  with a front mouth opening of
about 35~cm. At 3-4 places a halogen bulb was kept and normally one or two bulbs were alight at a time. The intensity of the light source is controlled by a constant current source. 
With this setup the sphere provides a highly uniform light over the opening of the sphere.
Above the opening mouth of the sphere a slider is kept that can carry the photographic plate.
This moves in the horizontal direction in and out of the sphere's mouth. The 
photographic plates were put on the slider whenever the digitization of the plates 
was carried out. A CCD camera was kept in the vertical slider.
The vertical slider helps in adjusting the size of the final image and also helps in focusing
with fine adjustments. An imaging lens with negligible vignetting and aberrations was kept
in front of the CCD camera. A green color broad-band filter was also kept in-front of the CCD
to reduce the intensity and heat load  on the CCD camera. A scientific grade CCD camera
with a 4k$\times$4k format and 16 bit readout at a rate of 0.5~MHz provided by Andor technologies. 
A 15$\mu$m pixel makes a 62~mm size CCD array. A 16 bit camera with a four port
read-out at a rate of 500~kHz provides images with high photometric accuracies and a wide dynamic
range. The CCD is cryogenic-cooled and operates at a temperature of --100$^{\circ}$C. With this 
temperature the CCD provides a low dark current and low readout noise. Two digitizer units 
with the same setup were used for digitizing the white-light images.

\section{Digitization of white-light images}
\label{sec:dwli}
The photographic plates are illuminated with uniform light source coming from the sphere. The images 
of the photographic plates were taken with an exposure time of 5-30 sec depending on 
the transparency of the plates. The higher the transparency, the shorter the time of exposure, and vice versa.
While digitizing the data sets, the dark current was subtracted from each image.
The flats were taken by placing a clean glass plate on the horizontal slider, which is
of the same thickness as the photographic 
plates. Before starting each month's data digitization, two to three flats were taken. The same setup 
and procedure was followed for  all data sets. 
Data were stored on the hard disk of the computer at Kodaikanal observatory and a copy
was kept at the Data Center of the Indian Institute of Astrophysics, Bengaluru. Thumbnail images of all raw and calibrated data are available through the website \url{http://kso.iiap.res.in}. Daily images are available through a search engine developed in-house at IIA using Java and Mysql database. The website shows raw and calibrated jpeg images starting from 1904 until today. The final data products along with full-resolution FITS images will be made available for the scientific community through this web portal in the near future. 

\begin{figure}[!h]
\begin{center}
\includegraphics[width=0.45\textwidth]{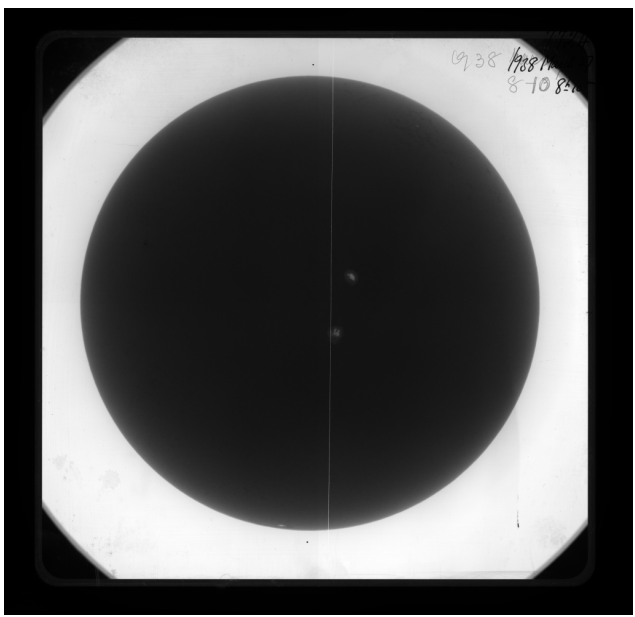} 
\end{center}
\caption{ Digitized white-light image of the full disk of the Sun. The image was digitized
with 4096$\times$4096 pixels resulting in a plate scale of 0.62 arcsec per pixel.
The central white line represents the east-west line on the sky. The date and time of the 
observations are shown in the top right corner of the image. The image was obtained on March 27, 1938 at 08:10~hr.}
\label{fig:2}
\end{figure}

Figure \ref{fig:2} shows a representative example of the digitized white-light image. The image is a photographic negative. A bright vertical line in the center of the image indicates the location of the east-west 
direction in the sky. The date and time of the observations are written in one corner of the image. 
Beyond the solar disk the unexposed part of the plate is white and can be used for 
stray-light estimation. Sometimes there are scratches and dust/fungus on the plates. The dust that is located
on the reverse side of the emulation was removed with a clean cloth. 
However, there still remain some dust particles  on the emulation side.

\section{Data calibration}
\label{sec:dc}
Before using the digitized data for scientific analysis, it is essential to calibrate  the data. This
includes the flat-fielding of the digitizing unit, centering the solar disk with the image disk,
orient the north polarity upward and correct for the photographic density values. We describe each 
of these in the next subsections.
\subsection{Flat-fielding of the digitized data}
For each month of data digitization, two to three flat-field images were recorded and saved. These flat 
images are dark-subtracted. The average flat 
-field was constructed and later used to correct the CCD pixel-to-pixel gain variations. 
This was done for all images at their original resolution.
\subsection{Locating the solar disk center and radius}
\label{sec:dfsdcr}
We locate the solar disk center and radius in terms of pixels or arc seconds. Many different methods have been adopted in the past \citep{denker1999}. We  
used a circle Hough transform to locate the solar disk center and radius \citep{ballard1981}. First we  
reduced the data to
one fourth of the original size. We then used a Sobel filter to isolate the sharp edges in the
solar images. This procedure not only detects the solar limb, it also detects the borders of the
features and sharp gradients. To separate the solar limb from the rest of the features we used
an intensity threshold of five times the mean value of the intensity of the Sobel-filtered solar disk.
This procedure clearly isolates the solar limb from the rest of the features. To use the 
circle Hough transform it is essential to feed the approximate radius of the intended image. 
We first fed the approximate radius of the reduced-size image to the circle Hough transform procedure. 
This procedure provides the center and radius of the solar disk. Four times the value of this
has been fed again into the circle Hough transform procedure to locate the center and radius of the
original solar white-light image. The limb co-ordinates of the original 
size Sobel-filtered image was introduced to search for the center and radius of the original image.
The circle Hough transform provides the center and radius of the solar disk with high accuracy. 

\subsection{Conversion to relative plate density}
\label{sec:cri}
In the exposed and developed photographic plates, the blackening corresponds to the incident flux
on the plates. The blackening also depends on the exposure time of the plate to the incident flux. The 
blackened area corresponds to the optical density of the exposed layer. This also suggests which
part of the plate is absorbed a light intensity. The photographic plate density is related to the 
logarithm of the plate exposure through the product of the incident light intensity and exposure time.
From the relation one can convert the plate density into the solar intensity values. However, there
are two difficulties  for incorporating this procedure into the calibration of the white-light images. 
The first one is that the density curves are available in the plates only after 1969.
The second  is  that the density curve changes from day to day. To overcome these problems
we calibrated the plates to  relative plate density rather than to absolute intensity values. 
For this we used available information in the plate with the 
formula \citep{mickaelian2007} $I = (V-B)/(T_{i} - B)$ , 
where I is the  relative plate density of the calibrated image in arbitrary units, $V$ is the 
average pixel value for the unexposed part of the plate, $B$ is the average value of the darkest pixel 
and $T_{i}$ is the density value at each pixel location. This procedure is similar to the one 
adopted by \cite{ermolli2009} to calibrate the Ca~II~K images obtained from various observatories 
around the world. 

\subsection{Image-centering and rotation}
\label{sec:icr}
Most of the time the solar disk in the photographic plate is not in the center of the image window.
From the procedure adopted in subsection \ref{sec:dfsdcr}, we have the information about the center and 
radius of the solar disk in the image window. With this information we shifted the solar disk 
so that the center of the image window and the center of the solar circle match. 
Later, we automatically detected the east-west line in the original image that goes from bottom to 
top in the image. This was made 
by selecting the large rectangular sized window in the center of the image and using a threshold value 
to detect the line automatically. We then computed the angle made by this line with respect to 
the vertical line. We then computed a P-angle for the epoch and added  the P-angle to  the
east-west 
angle and then rotated the image by this angle. This procedure  oriented the images in 
such away that the northpole of the image is always directed upward. Later,
we examined each and every image for the north polarity direction to make sure that it was
performed uniformly in all images.

\begin{figure}[!h]
\begin{center}
\includegraphics[width=0.45\textwidth]{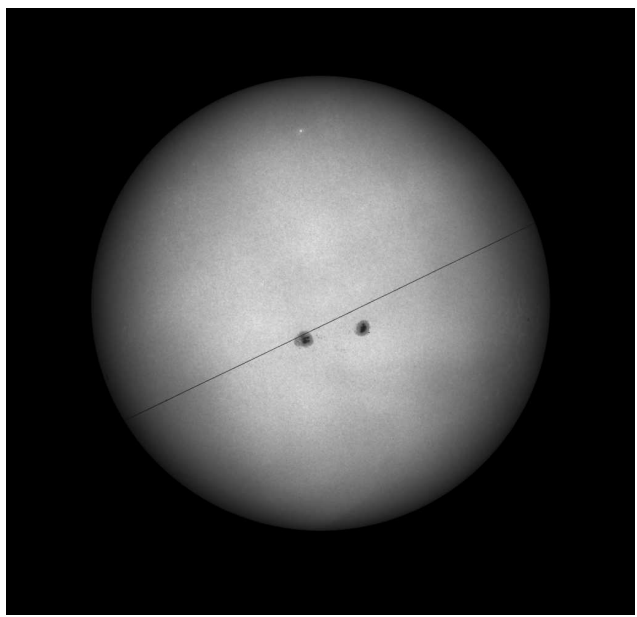} \\
\end{center}
\caption{Calibrated and aligned white-light image.  The full image size
is 4096$\times$4096 pixels. Each pixel corresponds to 0.62 arcsec. The image was obtained on 
March 27, 1938 at 08:10~hr.}
\label{fig:3}
\end{figure}

Figure \ref{fig:3} shows the processed image sample of calibrated and aligned white-light image. 
Clearly, the sunspots are aligned in the east-west direction. In many of the images the 
boundary between the umbra and penumbra can be clearly demarcated. Even the limb has a sharp 
boundary in the image. Apart from these there is some dirt in the image that is seen as white 
or black spots depending on whether it is dirt or a scratch on the image. The dark line is the east--west line
on the sky.

\subsection{Corrected image and image header}
\label{sec:ciih}
In the final stage of the calibration procedure we saved the corrected images (which includes all 
procedures  mentioned above) in the FITS format with relevant information in the header. This
includes the date and time of the photographic plate, the center and radius of 
the solar disk in the image, and the approximate pixel resolution of the image. 

We calibrated the white-light data starting from 1920 to 2011, which covers a period of 
about 92 years. But 15 years of data still have to be calibrated. In this period, about four years of
data have a different spatial resolution. The solar image is much larger than the images available after
1920. The images taken from 1905 to 1909 have north-south and east-west cross lines. In the second 
stage of calibration we discuss these issues. 

\section{Properties of white-light image time series}
The 106 years of white-light data sets are not uniform throughout the time series. Each day image 
is different from the next in contrast. This is because the seeing is
different on different days of the observations and sometimes passing clouds were present
when the images were taken. Moreover, the image acquisition has changed from photographic plates to films
in the later part of the twentieth century. Although the optics was not changed over 90 years, there could be
degradation in the quality. There could be an increased scattered light in the images. This eventually
degrades the quality of the images. Some of the properties of long time series of the data
sets are compared below. Some of these procedures are similar to those followed by 
\cite{ermolli2009}.

\subsection{Image contrast}
\label{sec:ic}
The contrast of the image is a quantity that provides information about the quality of the image.
 The images are not intensity-calibrated, which requires the plate $\gamma$ parameter, which 
depends on the plate emulsion and the plate development process \citep{devauc1968}. The plate 
$\gamma$ parameter 
is an indicator of the image contrast. Naturally, the image contrast measured over many years
indicates how the plate $\gamma$ parameter changed over the years.
We measured the variation in the relative plate density over the solar disk from center to  
limb. This
was made by constructing 25 rings of equal area from center to limb. At each ring we 
obtained a median value. From the set of median values we obtained the minimum and 
maximum values.  We then computed the contrast of the image by taking the ratio of the difference 
and sum of the maximum and minimum values.

\begin{figure}[!h]
\begin{center}
\includegraphics[width=0.45\textwidth]{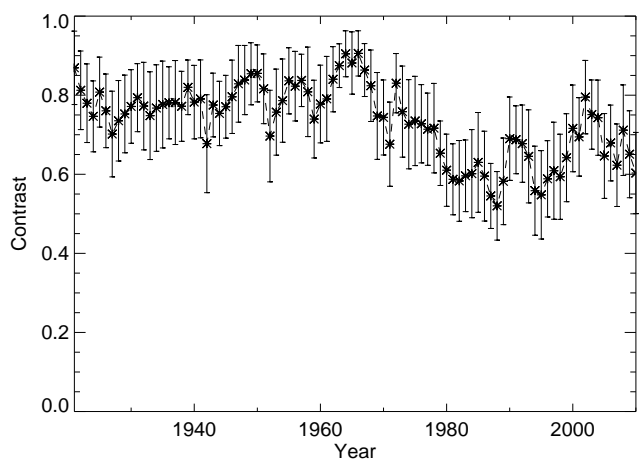} 
\end{center}
\caption{Yearly averaged image contrast as a function of time. The vertical bars 
represent the root mean square value of the contrast computed over the year of white-light data.}
\label{fig:4}
\end{figure}

Figure \ref{fig:4} shows a plot of the mean value of the image contrast over the 90 years of data. 
The contrast values were obtained by averaging the data values over the years. The image 
contrast was varying from 0.8-0.9 starting from 1920 to 1970. However, there is a decline 
in the image contrast from 1970 onward until the recent past. This could be due to  
various reasons. The optics were possibly old and were maybe not cleaned over the time. The photographic
plates have changed to films during this period. It could also be due to the seeing, which has 
changed over the time. By looking at the data over all the years there is a definite change in the image
clarity from 1920 to 2011. The vertical bars represent the rms value of the contrast over the year.

\subsection{Spatial resolution}
\label{sec:sr}

\begin{figure}[!h]
\begin{center}
\includegraphics[width=0.45\textwidth]{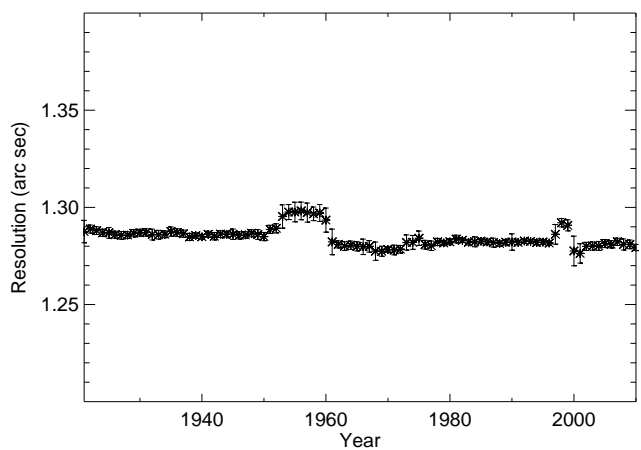} 
\end{center}
\caption{Yearly averaged spatial resolution of the white-light image as a function of 
time. The vertical bars show the root mean square value over the data points. In many cases 
it is lower than the width of the data points.}
\label{fig:5}
\end{figure}

The spatial resolution of the image provides an idea about the spatial information in the images. 
The central 
128$\times$128 box was used for estimating the spatial resolution of the white-light images. 
A power spectral analysis was adopted to compute this parameter. A majority of the 
information about the features observed in the images lies in the low-frequency range. The frequency at
which 98\% of the total power spectral density remains is taken as the spatial resolution of the
images. Figure \ref{fig:5} shows the variations of the spatial resolution over a period of 90 years.
The yearly average of spatial resolution is plotted here. The vertical bar represents the 
rms value. The rms values are very low, they have the same size as the width of the symbol
used in the plot. During the period from 1956 to 1960 the spatial resolution is lower than
rest of the data points. However, the difference is very small.
From the plot it is clear that the spatial resolution is almost constant over time with small variations. 

\subsection{Ellipticity of the image}
\label{sec:eti}

\begin{figure}[!h]
\begin{center}
\includegraphics[width=0.45\textwidth]{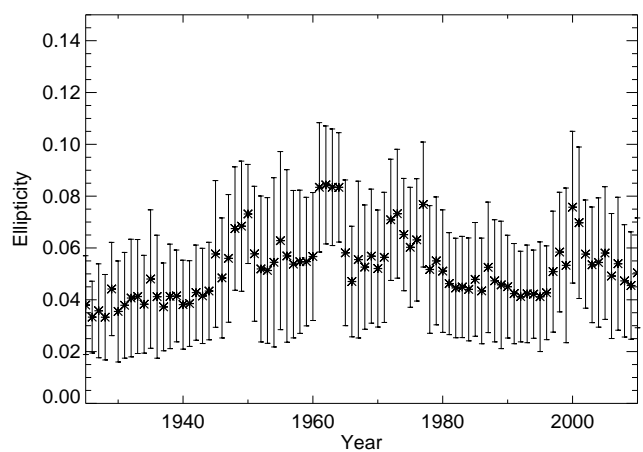}  \\
\end{center}
\caption{ Temporal variations of the image shape. The data points are 
obtained from yearly averaged values.}
\label{fig:6}
\end{figure}

\begin{figure}[!h]
\begin{center}
\includegraphics[width=0.45\textwidth]{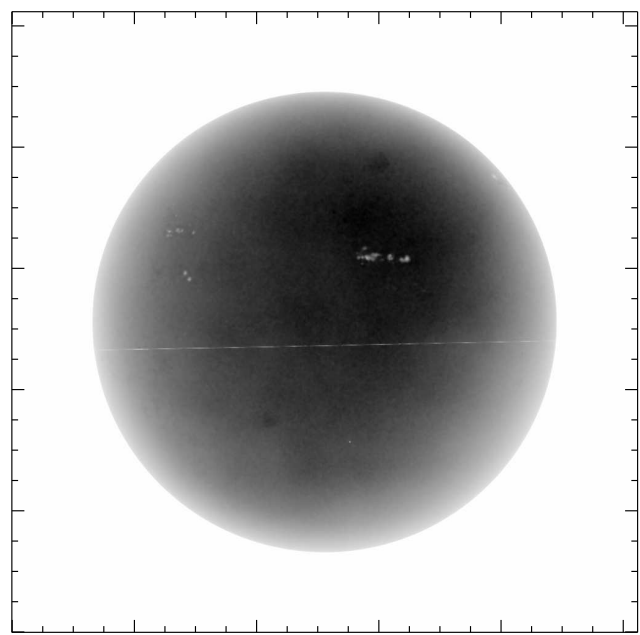} 
\end{center}
\caption{Inverted image used for detecting the sunspots. The image size is 1024$\times$1024 pixels,
each pixel corresponds to 2.5 arcsec. The sunspots are visible in the full-disk image as white spots.
The image was obtained on January 08, 1938 at 08:18~hr.}
\label{fig:7}
\end{figure}
Most of the images taken earlier in certain wavelengths are not perfectly circular in shape. This is
true for the images made using the scanning spectrometer. However, the direct images taken from 
the telescope are not distorted. They appear almost circular in shape unless the photographic
plates were kept oblique to the image. To determine the ellipticity in the 
white-light images we first detected the limb in the original image with the Sobel filter. The 
ellipse-fitting procedure provided the length of the two axes. Then the eccentricity was 
computed as $eccentricity = \sqrt{1-(r_{minor}/r_{major})^{2}}$, where $r_{major}$ and $r_{minor}$ are 
the length of the major and minor axes. The ellipticity was computed for 90 years
of data.

Figure \ref{fig:6} shows a plot of the ellipticity variation in the image over time.  
Each point shows the yearly eccentricity average and the vertical bar is the rms value 
computed over the year. The plot shows that there is a slight variation in the ellipticity, which could
be due to the quality of the image, or to a slight variation in fitting the elliptical to the
detected limb points. The ellipticity value is also very low and is close to the 
ellipticity value of the recent data taken with the CCD cameras \citep{ermolli2009}. We would like to point out
here that we did not measure the ellipticity that can arise from the atmospheric refraction. 
Previously, \cite{sivaraman1993} corrected for the atmospheric refraction in the 
limb and for the individual sunspot umbral area. In a subsequent study, we are planning to 
examine the effects of  
atmospheric refraction on the geometry of the image.

\section{Sunspot area coverage}
\label{sec:sac}
In white-light images the sunspots are the prominent structures visible on the solar disk. However,
near the limb side faculaes are visible as well. The sunspot number and the area coverage vary 
over the solar cycle and also from cycle to cycle.  A strong solar activity is visible on the 
sun whenever many sunspots with larger area coverage appear on the solar disk. On the
other hand, the solar activity reaches minimum when there are fewer sunspots.
Though the sunspot areas were analyzed in the past with various data sets, it would be 
interesting to examine the sunspot areas obtained from the Kodaikanal white-light images with
modern techniques of feature detection. 

To identify the sunspots in white-light calibrated images we used a modified version
of the sunspot tracking and recognition algorithm (STARA: \cite{watson2009}). First, we reduced
the size of the data from 4k$\times$4k to 1k$\times$1k. Then the image was inverted 
in such a way that the sunspots appear as bright as shown in Figure \ref{fig:7}. Following 
\cite{watson2009}, we used a top-hat transform on the inverted Kodaikanal white-light images.
The top-hat transform is an original image that is subtracted from the image, which 
is modified by
the opening operation. This procedure makes the background flat and the sunspots appear as peaks. More
details on this algorithm can be found in \cite{watson2009}. A slight difference between their
method and ours is that we applied this procedure twice to the digitized white-light images 
with different intensity thresholds. In both cases we used same structuring element with a 
circle of diameter  of 16 pixels. In the first step it detects all sunspots that appear near the 
limb and close to the limb. In the second run it identifies all sunspots that appear inside 
the solar disk up to $\mu = 0.2$. Although the sunspot detection is automatic, the procedure 
not only detects the sunspots, it also detects the dirt and the east-west line in the sky. To 
remove all these additional features we added contours on the detected features and then used a cursor to 
identify the contours that belong to the sunspot. Despite this is slight human intervention,
 this provides the best results in discarding the redundant features on the photographic plates. 

\begin{figure*}[!ht]
\begin{center}
\includegraphics[width=0.45\textwidth]{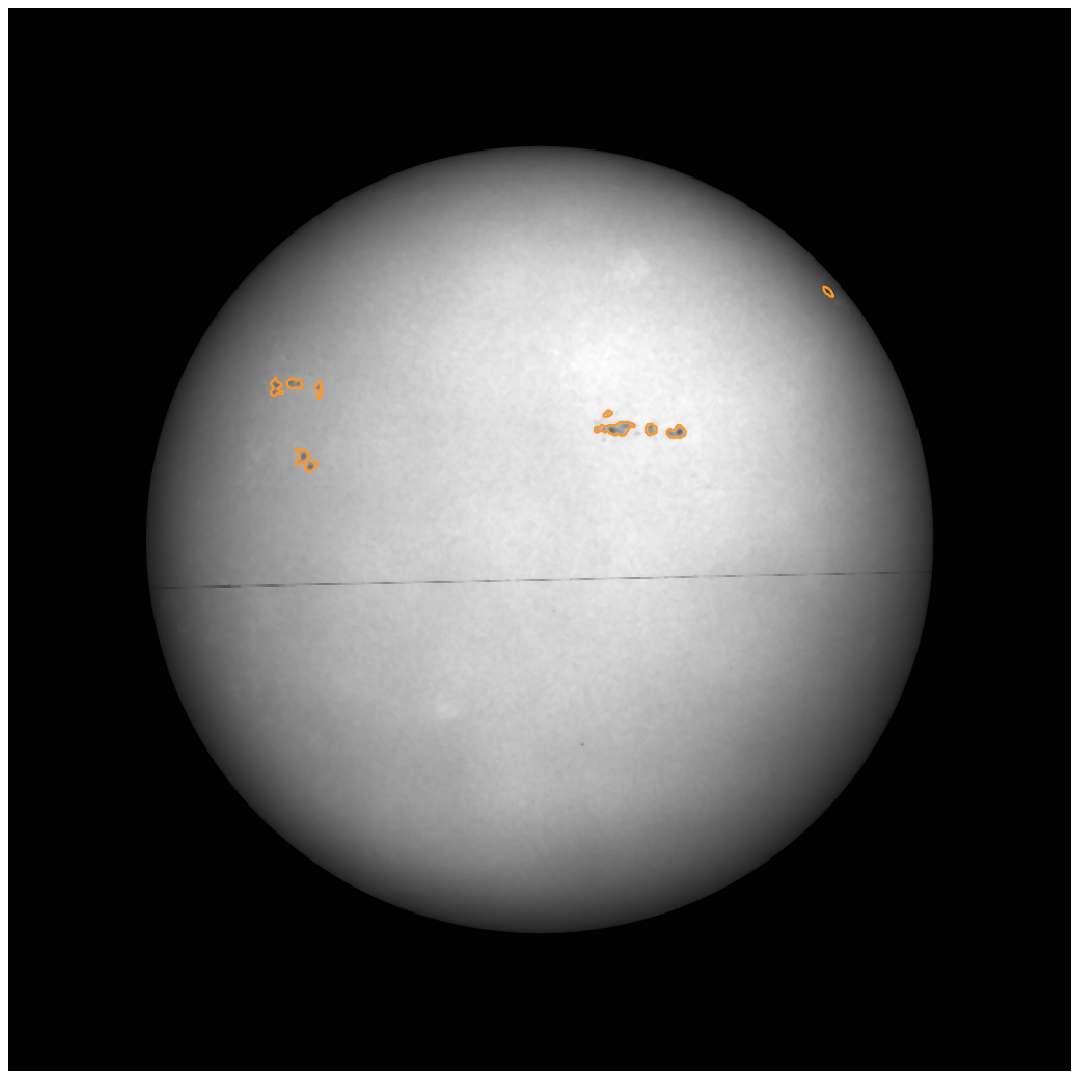}\includegraphics[width=0.45\textwidth]{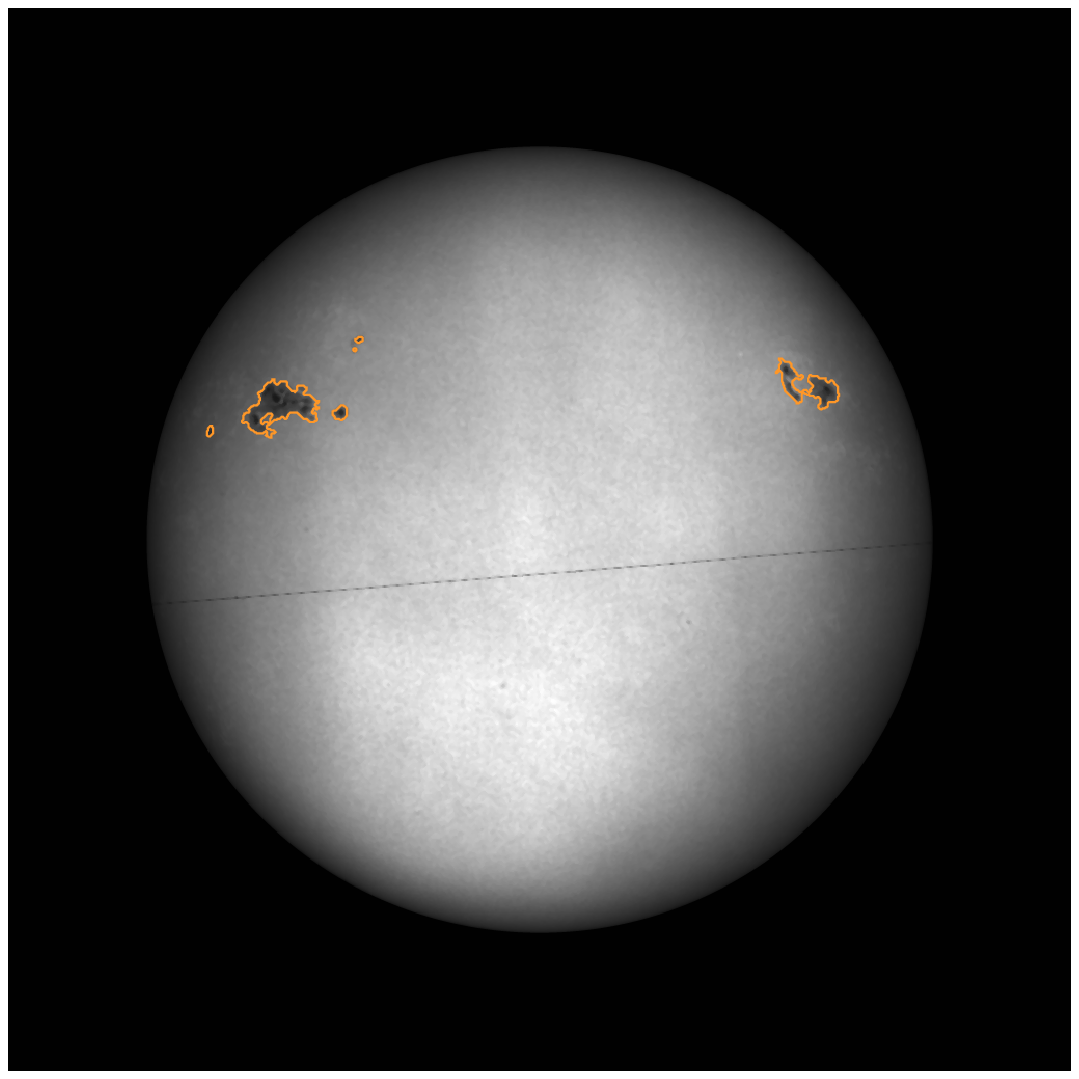}  \\
\end{center}
\caption{Sunspots with the contours overlaid upon the image. Left and right side
panels correspond to images recorded on Jan 08 and Jan 15, 1938.
The image size is 1024$\times$1024 pixels and each pixel corresponds to 2.5~arcsec.}
\label{fig:8}
\end{figure*}

Figure \ref{fig:7} shows the inverted calibrated image. There are several sunspot  groups
visible on the disk. Some of the spots are huge and others are small. 
Figure \ref{fig:8} (left panel) shows the sunspots with the contours. 
Here the sunspot group is small and the algorithm detects all small sunspot groups.
Figure \ref{fig:8} (right panel) shows another example. The large sunspot group is detected 
by the algorithm. The
area of the detected sunspot group is corrected for the area foreshortening, then 
the sunspot area is converted into the millionth of a hemisphere on the Sun. Finally, we summed the 
entire sunspot area that appeared on the Sun on that day and stored it in a file where the first 
column indicates the date of the observation and the second column indicates the sunspot area. 
This method of estimating the sunspot area does not distinguish between the northern and 
southern hemisphere. 

\begin{figure*}[!ht]
\begin{center}
\includegraphics[width=0.45\textwidth]{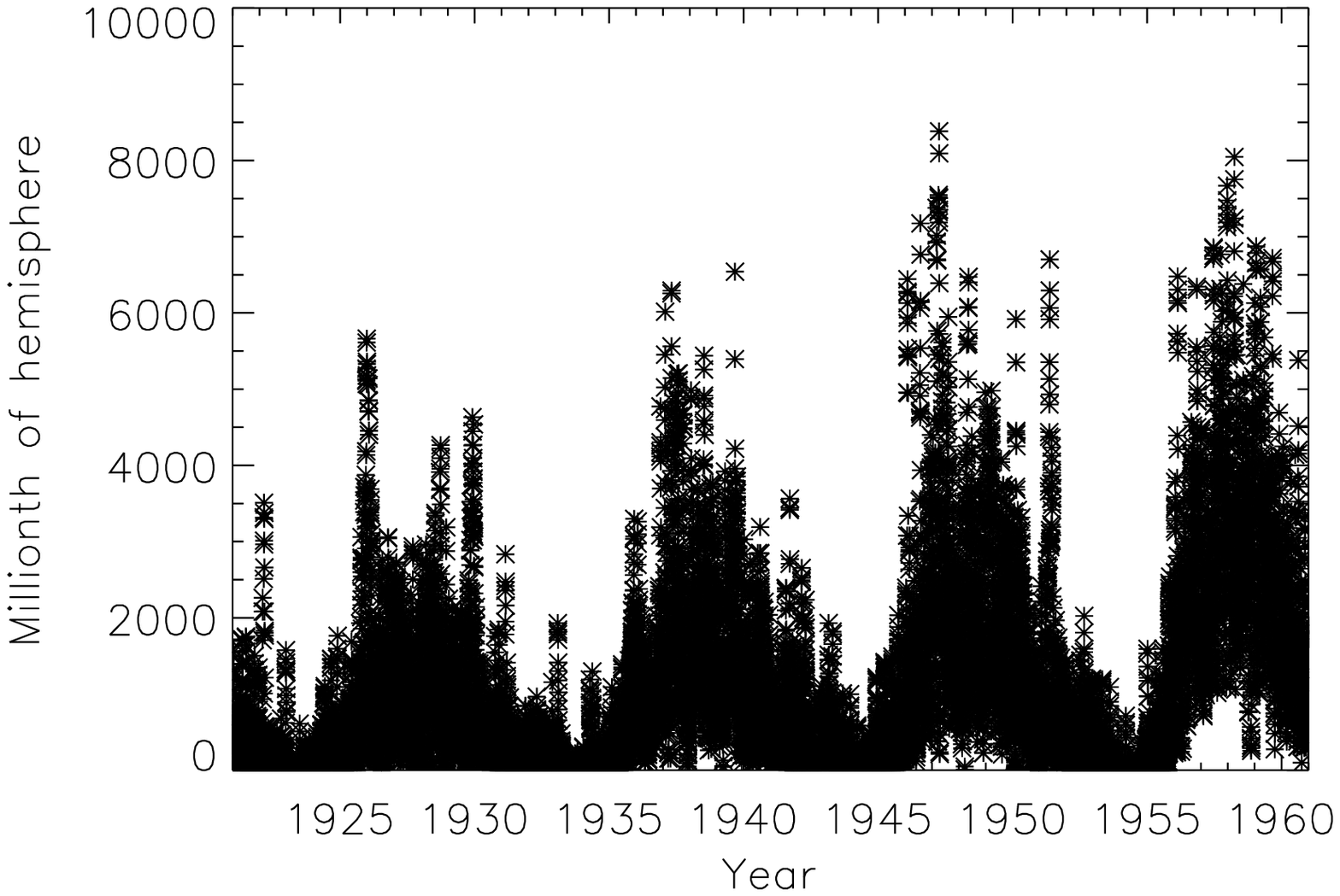}\includegraphics[width=0.45\textwidth]{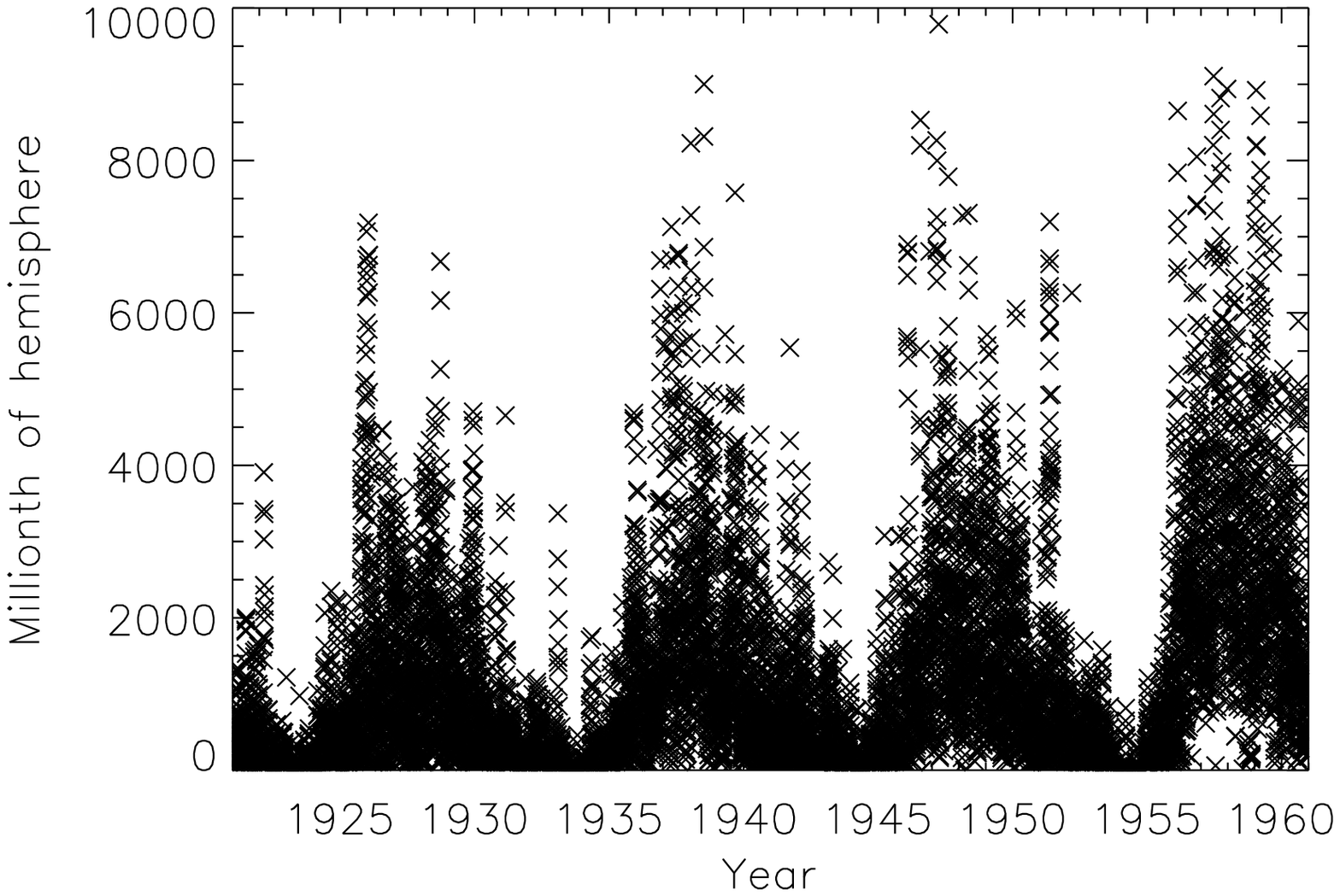} 
\end{center}
\caption{Left: Daily sunspot area computed using the Kodaikanal white-light data as a 
function of time. Right: Same as the left-side plot but for Greenwich data.}
\label{fig:9}
\end{figure*}

A modified version of STARA was applied to the calibrated white-light data to detect the sunspots,
i.e.,  from 1921 to 1960 spanning about 40~years. The data set includes four solar cycles. 
Figure \ref{fig:9} (left panel) shows a plot of the daily sunspot area shown over 40~years. The plot clearly
shows the solar cycle variation of sunspot area. During the maximum, the sunspot area coverage over
the solar disk exceeds an  8000 millionth of hemisphere ($\mu$H). However, the area coverage  
varies for different solar cycles. In solar cycle 16 the area coverage is small. On the other hand, 
cycle 19 has a large area coverage compared to cycle 16. We also compared our  sunspot area results 
with the Greenwich sunspot area compiled by David Hathaway 
(\url{http://solarscience.msfc.nasa.gov/greenwch.shtml}) . Figure \ref{fig:9} (right panel) shows the 
plot of the Greenwich sunspot area for the same period. Comparison of this
plot with Figure \ref{fig:9} (left panel) shows the similarity between the two except for the maximum
area coverage during the sunspot maximum. Otherwise the pattern of the area coverage is almost
the same, which suggests the goodness of the data and method of identifying the sunspots and groups.

\begin{figure*}[!htbp]
\begin{center}
\includegraphics[width=0.45\textwidth]{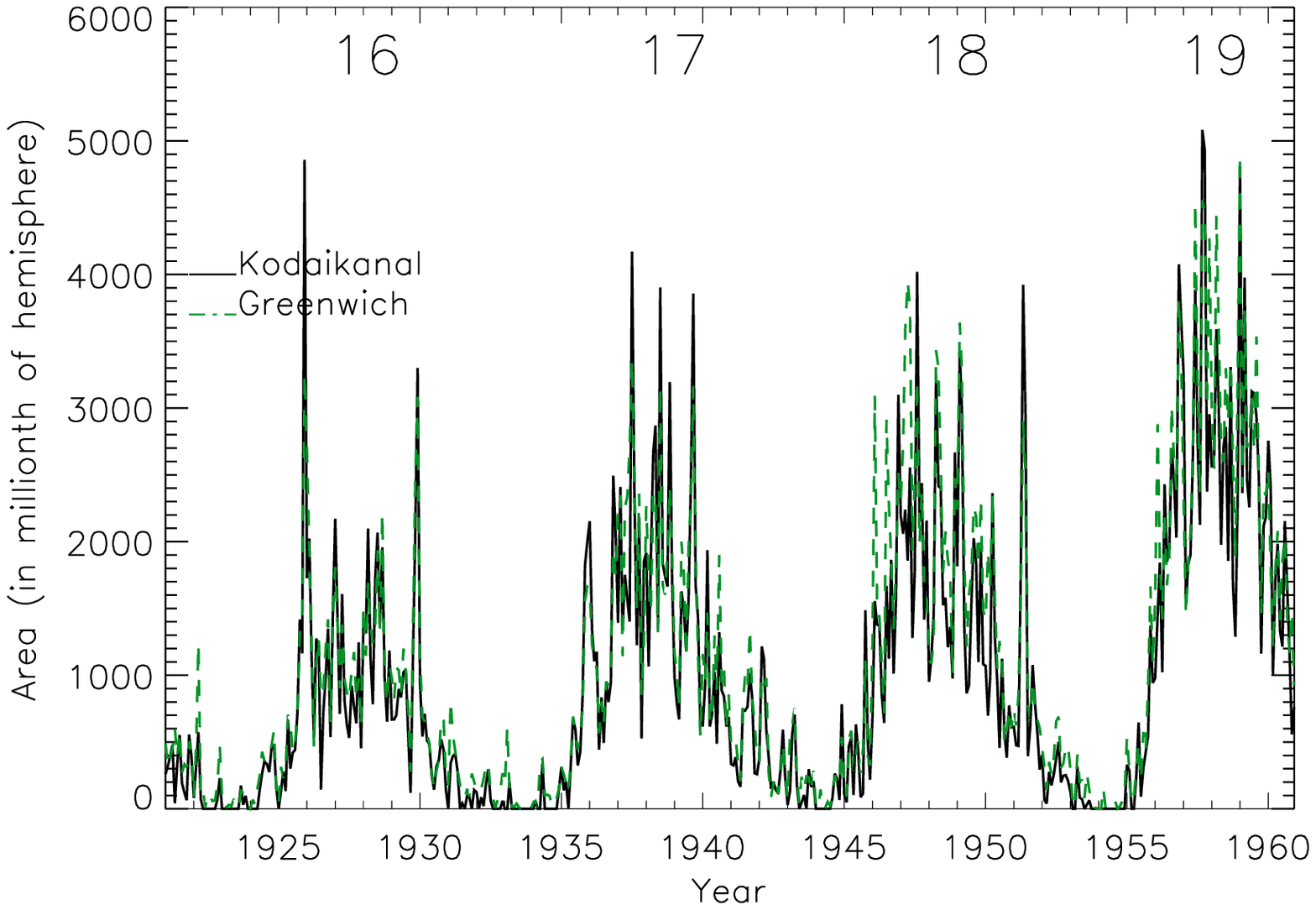}\includegraphics[width=0.45\textwidth]{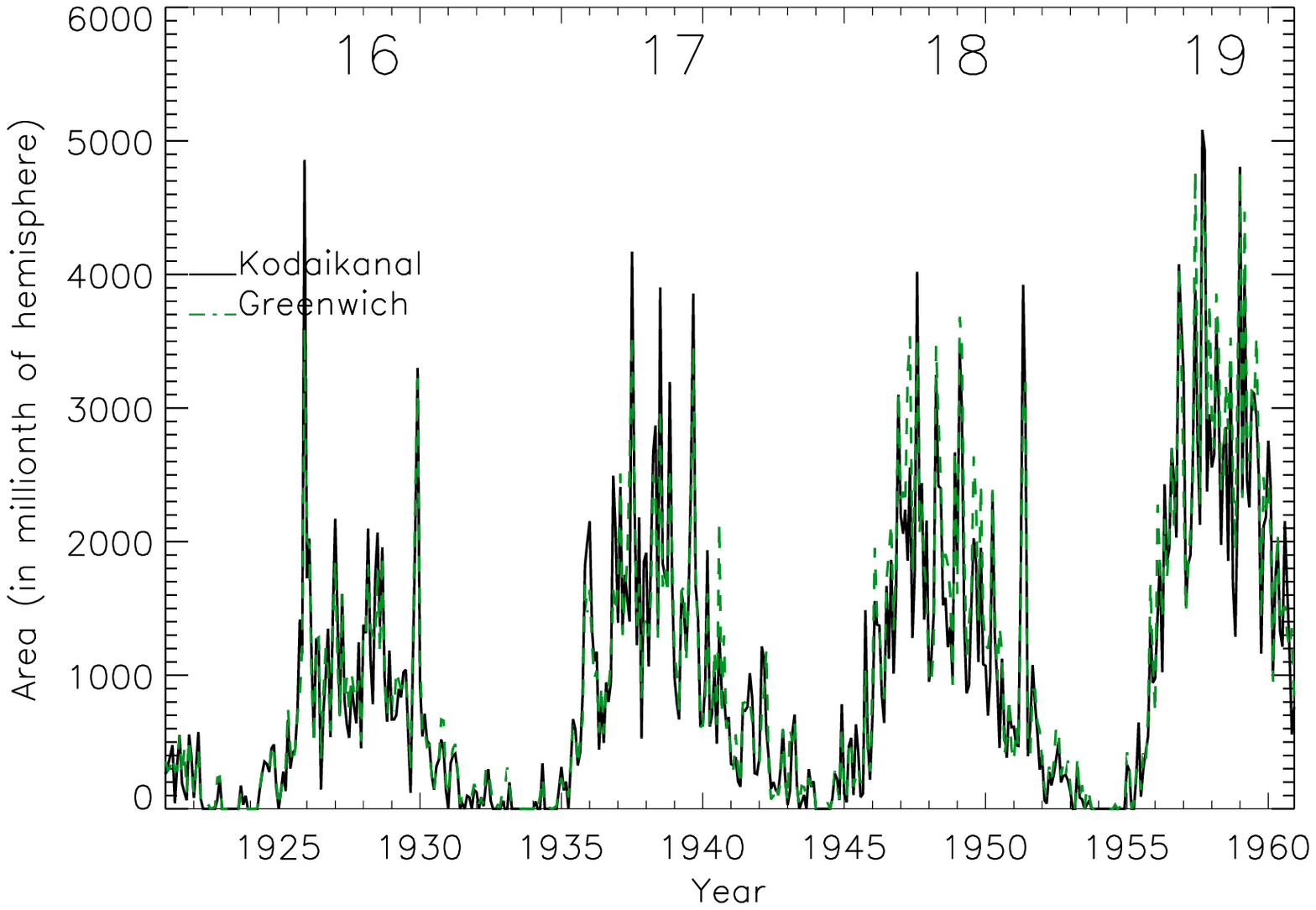} \\
\end{center}
\caption{Left: Monthly averaged sunspot area data obtained from Kodaikanal and Greenwich are overplotted on each other. Right: Same as the left-side plot but for the monthly median.}
\label{fig:10}
\end{figure*}

Figure \ref{fig:10} (left panel) shows a monthly average of the Kodaikanal sunspot area (thick dark line) 
and the monthly average of the Greenwich sunspot area (dashed line with green color). The plot shows 
that the two monthly averaged sunspot
areas match very well with a correlation coefficient of 0.97. Most of the time the Kodaikanal 
monthly average is little higher than the Greenwich monthly averaged sunspot area. However,
the opposite is also true. We also plotted the monthly median of the sunspot area
calculated from both  data sets (Figure \ref{fig:10} (right)). Here we see a similar correlation 
between the two data sets. 

We then compared the yearly averaged sunspot area of the Kodaikanal and Greenwich data 
(Figure \ref{fig:11}).
Although it appears that the pattern is almost the same, there are certain differences 
between them.  In some cases the year of the sunspot maximum is different.  
At other  times there is a double peak in the Kodaikanal data that is absent from 
Greenwich data. Figure \ref{fig:12} shows the scatter plot of the yearly averaged sunspot area
of the Kodaikanal and Greenwich data sets. The obtained correlation coefficient is 0.99.
The scatter is very small in the smaller area and it is larger for the larger area. 

In Figure \ref{fig:13} we compare the monthly averaged Kodaikanal sunspot area with the monthly 
averaged Greenwich sunspot number. The comparison shows that the correlation between the 
sunspot number and the sunspot area is good. However, there are certain differences between them. 
For example, in cycle 17 during the rising phase of the sunspot number there is a sharp rise
in the sunspot area, but the increase in sunspot number is marginal. These differences can be found
in many places. On the other hand, although there is an increase in 
 sunspot number in cycle 19 during the peak time, the sunspot area decreased, meaning there could be many 
small sunspots whose total area will be small. This kind of anti-correlation between sunspot number and
area can be found in many cycles. It suggests that there could be a large sunspot with larger apparent area although the number of 
sunspots  on the Sun could be small. This can be clearly seen in the scatter plot of the monthly averaged 
Greenwich sunspot numbers and the monthly averaged Kodaikanal sunspot area (Figure \ref{fig:14}). 
In the plot at some places there are few spots, but the area is large. 
\begin{figure}[!htbp]
\begin{center}
\includegraphics[width=0.45\textwidth]{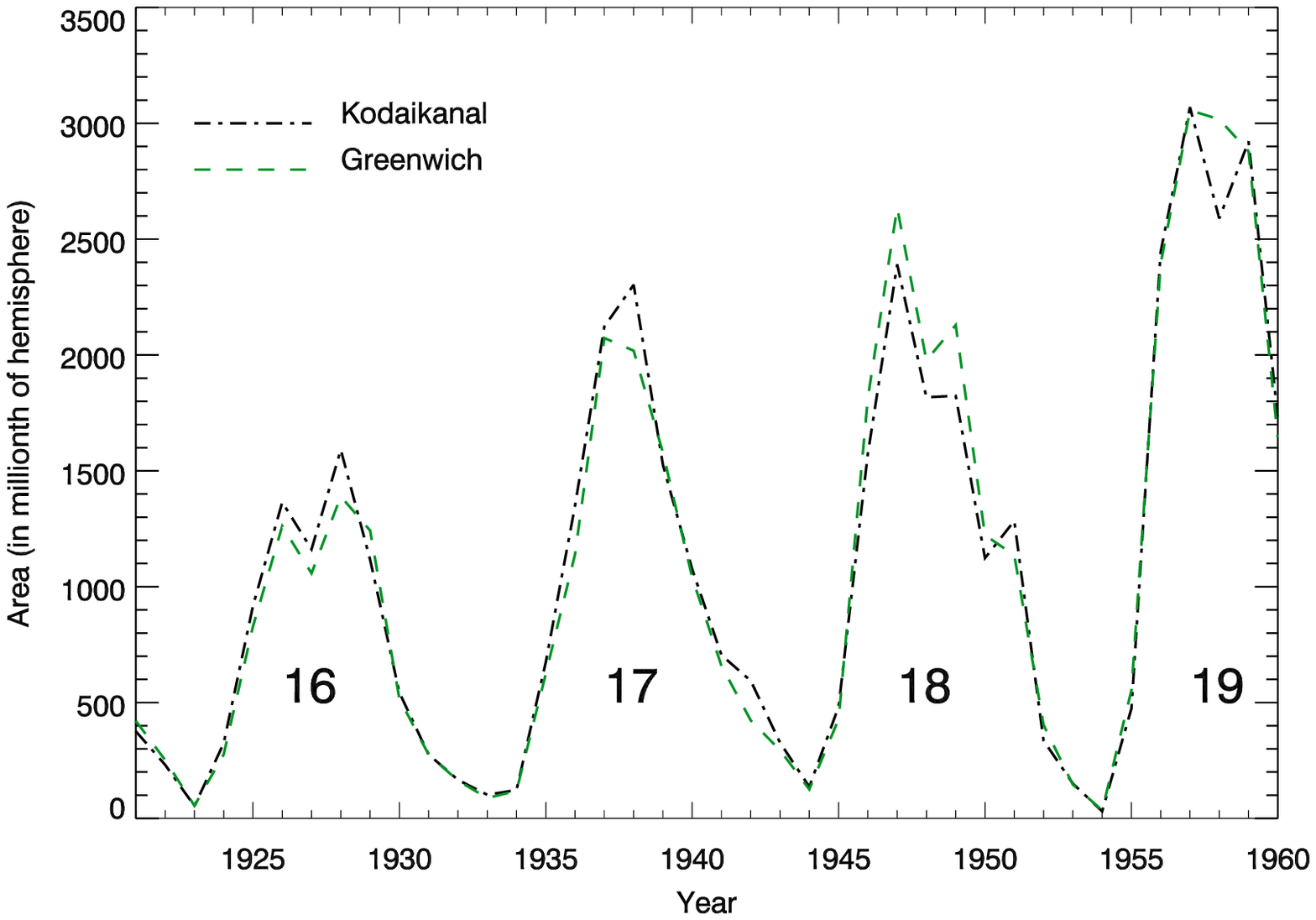} 
\end{center}
\caption{Yearly averaged sunspot area data obtained from Kodaikanal and Greenwich overplotted.}
\label{fig:11}
\end{figure}
\begin{figure}[!h]
\begin{center}
\includegraphics[width=0.45\textwidth]{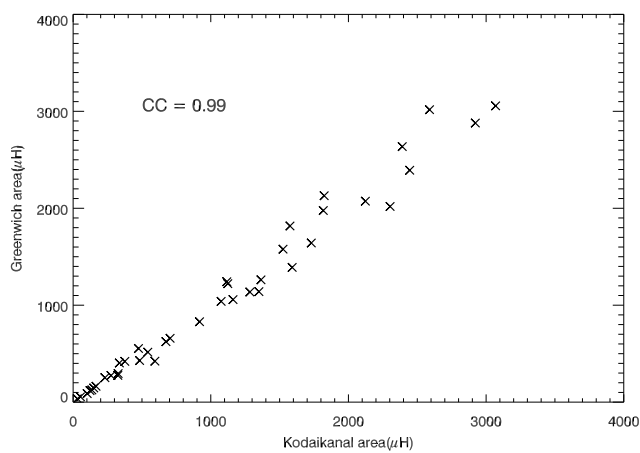} 
\end{center}
\caption{Scatter plot of the Kodaikanal and Greenwich sunspot area.}
\label{fig:12}
\end{figure}
\begin{figure}[!h]
\begin{center}
\includegraphics[width=0.45\textwidth]{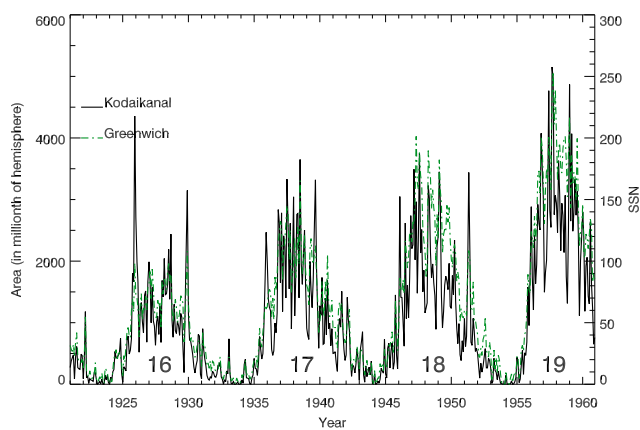} 
\end{center}
\caption{Monthly averaged sunspot area obtained from Kodaikanal and the sunspot number obtained from 
Greenwich for four solar cycles.}
\label{fig:13}
\end{figure}
\begin{figure}[!h]
\begin{center}
\includegraphics[width=0.45\textwidth]{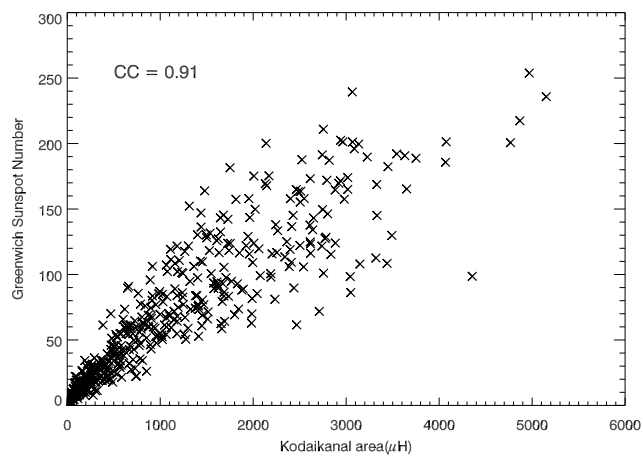} 
\end{center}
\caption{Scatter plot of the monthly averaged sunspot area and sunspot number.}
\label{fig:14}
\end{figure}

\section{Conclusions} 
\label{sec:sd}
In Kodaikanal, white-light observations of the Sun on  photographic plates 
and films have been made since 1904 . The white-light images have been collected for 
more than 105 years with 290 images on average per year. These data sets were recently digitized
at Kodaikanal and about 90 years of data were calibrated and aligned. They have a
uniform spatial resolution. This is mainly because the same optics were used throughout 
the 90 years, starting from 1920 until today. Before 1920 there were some changes to the optics
and hence the obtained images are slightly larger. The contrast of the images is quite good until
1970, after which it started to decrease slightly. This is the period where the observations changed from 
photographic plate to films.  In the linear part of the characteristic curve, the density
parameter of the plate is proportional to the logarithm of the exposure parameter with the  $\gamma$ function
as a slope. The $\gamma$ parameter depends on the emulsion of the photographic plates and also
on how the plates were developed. In a way the $\gamma$ parameter represents the image contrast. The image 
contrast declined after 1970, suggesting that there is a change in the $\gamma$ parameter, which likely
implies a change in the plate emulation composition. This also indicates that for
a detailed quantitative analysis one needs to use the $\gamma$ parameter for intensity calibration
purposes. We intend to do this in the future. But for the current study of the sunspot area this may not
be required.
The image shape is almost circular and does 
not show any oblateness. We performed a relative calibration to the digitized images 
by using the information available on the plates. We also corrected for the image 
orientation and centering. However, we did not yet correct for the scattered light, which we plan for 
the next step.

Sunspot area measurements over a long time period are
valuable and important data for studying the solar activity on the Sun. The solar activity is responsible 
for the variations in EUV and X-ray radiation emitted from the Sun. The changes in the 
Earth's upper atmosphere are modulated by the observed solar activity \citep{she2002}. The 
digitized Kodaikanal white-light images could be complimentary data for studying the variations in 
sunspot and faculae areas over 100 years that may have influenced changes in the Earth's upper atmosphere. 
 By using a semi-automated code to detect the features on the Sun we 
identified and detected all sunspots visible on the Sun. We measured 
sunspot area and groups of sunspots on each day. A correction for the projection effect was
also included. While the measured area is  well-correlated with the Greenwich sunspot 
area, the area measured by the Kodaikanal white-light data is slightly larger than that of Greenwich.
There is a  high correlation of monthly mean and median 
values of the sunspot area of the Kodaikanal and Greenwich data. However, there 
is a slight offset between the peak times of the sunspot maximum in a few cycles when we
looked at the yearly averaged sunspot area of Kodaikanal and Greenwich. The yearly averaged
sunspot area of Kodaikanal and Greenwich is very highly correlated at 0.99. Even the monthly averaged 
sunspot number obtained from Greenwich follows the pattern of monthly averaged Kodaikanal 
sunspot areas very closely. The scatter plot of the two shows a correlation coefficient of about 0.91.
This high degree of correlation of the Kodaikanal sunspot area with that of Greenwich 
suggests that one can also accurately extract the information about the features with the  Kodaikanal data . Hence, the sunspot area obtained from the Kodaikanal observatory white-light 
images can be directly used to estimate solar irradiance variations with a minimal intercalibration
with other data sets \citep{balmaceda2009}. 

Kodaikanal has been taking white-light, Ca K, and H$_{\alpha}$ data for the past 100 years.
Although many observatories have made similar kinds of observations in the past and in 
recent times, the Kodiakanal data are uniform over the past 100 years. The optics and processing of  the
plates were kept uniform. However, the photographic plate has been
replaced with films and at the same time there is a reduction in the contrast of the images. Hence,
to keep the quality of the data uniform, it is essential to calibrate the images to intensity format.
But for the current study of sunspot area calculations it is not required to  
intercalibrate the data. This is an advantage in studying the solar irradiance etc. 
(see also \cite{ermolli2009}).
A few more years of data need to be calibrated, which will be taken up soon. In the future, we aim to 
compare the results of  extracted parameters using different techniques with the
other data sets available all around the world. Our impression is that sunspot areas  may turn out to be a better proxy than sunspot numbers for studying the magnetic activity variation, particularly on shorter time scales.

\begin{acknowledgements}
We would like to thank  many observers at Kodaikanal,  who have observed the Sun in 
white-light and made it available to us. Special thanks go to Jagdev Singh, who has initiated this new digitization program several years ago. We thank  Rekhesh Mohan, N. Sneha, Nimesh Sinha, Ayeha Banu, 
S. Kamesh, P. Manikantan, and Janani for their help at different phases. We also would like  to thank 
the staff members at Kodaikanal, who  helped us in setting up the digitizer unit and carry out 
the digitization at Kodaikanal.  Finally we would like to thank the referee, R. Ulrich, for his valuable comments and suggestions, which improved the quality of the presentation.
\end{acknowledgements}

\bibliographystyle{aa} % style aa.bst
\bibliography{reference} % your references Yourfile.bib

\end{document}